\documentclass[11pt,letterpaper]{article}
\usepackage[english]{babel}
\usepackage{amsmath}
\usepackage{amssymb}
\usepackage{amsthm}
\usepackage{amsfonts}
\usepackage{graphicx}

\newtheorem{theorem}{Theorem}
\newtheorem{lemma}{Lemma}

\newtheorem{corollary}{Corollary}
\input{tcilatex}

\begin{document}

\title{On the Quantum Circuit Complexity Equivalence}
\author{Milos Drezgic \thinspace \thinspace\ \ \ \ \thinspace \thinspace
\thinspace\ Shankar Sastry \\
Department of Electrical Engineering and Computer Science, UC Berkeley}
\maketitle

\begin{abstract}
Nielsen \cite{Nielsen05} recently asked the following question: "What is the
minimal size quantum circuit required to exactly implement a specified $%
\mathit{n}$-qubit unitary operation $U$, without the use of ancilla qubits?"
Nielsen was able to prove that a lower bound on the minimal size circuit is
provided by the length of the geodesic between the identity $I$ and $U$,
where the length is defined by a suitable Finsler metric on $SU(2^{n})$. We
prove that the minimum circuit size that simulates $U$ is in linear relation
with the geodesic length and simulation parameters, for the given Finsler
structure $F$. As a corollary we prove the highest lower bound of\ $O(\frac{%
n^{4}}{p}d_{F_{p}}^{2}(I,U)L_{F_{p}}(I,\tilde{U}))\,$and the lowest upper
bound of $\Omega (n^{4}d_{F_{p}}^{3}(I,U))$, for the standard simulation
technique. Therefore, our results show that by standard simulation one can
not expect a better then $n^{2}$ times improvement in the upper bound over
the result from Nielsen, Dowling, Gu and Doherty \cite{Nielsen06}. Moreover,
our equivalence result can be applied to the arbitrary path on the manifold
including the one that is generated adiabatically.
\end{abstract}

\section{Introduction}

Quantum computation is inherently a process of continuous evolution of
quantum states that has the potential to fundamentally change the notion of
feasibly tractable computation. Only recently did researchers start to think
how notions from the differential geometry \cite{Warner} can be used to
represent this process. Instead, quantum circuits, as an inherently discrete
notion of computation, have been most commonly used to represent this
continuous process. Any quantum operation, a unitary matrix $U\in SU(2^{n}),$
is an element of a Lie group, and a point on the $\mathcal{U}\equiv
SU(2^{n}) $ manifold, whose tangent bundle can be endowed with the Finsler
structure $F $, that effectively provides a measure of length for any path
on the Finsler manifold $(\mathcal{U},F)$. In particular, the paths that we
are interested in are geodesics. These are locally and, under certain
conditions, globally minimal length paths between any two points on the
manifold. They are of particular interest, because if closely simulated they
can give the smallest circuit complexity for the given unitary $U.$

The aim in the approach that we take here is to tackle the question about
the complexity of the circuit necessary for the simulation of an arbitrary
unitary gate. As the length of the geodesic for the particular unitary is
its intrinsic property, ideally one would succeed in finding the minimum
number of circuits necessary to implement the unitary by simulating exactly
its geodesic. Therefore, the hope here is to learn about the circuit lower
bounds by basically transforming the hard combinatorial optimization
problems over large sets to the problems in continuous domain that can be
solved with tools of differential geometry and the calculus of variations.

One of the first results that had the flavor of this transform was
introduced by Mochon \cite{Mochon06}, who proved that in the discrete model
and the analogous continuous model, i.e. the Hamiltonian oracle model,
Oracle interrogation, the problem of computing XOR and Grover search have
the same complexity. Moreover, Nielsen \cite{Nielsen05} and subsequently
Nielsen, Dowling, Gu and Doherty \cite{Nielsen06} proved that for
particularly chosen metric there is a polynomial equivalence between the
geodesic length and number of gates necessary for the simulation. The lower
bound for the minimum number of gates necessary for the simulation has been
determined for exact simulation and the upper bound has been determined for
the arbitrary precision. The metric chosen in \cite{Nielsen06} penalizes all
those directions on the manifold that are not easily simulated by local
gates, so that coefficients for stabilizer elements of Hamming weight
greater than two bear high cost, i.e. have longer paths.

In this paper we prove the stronger result and show the exact upper and
lower bound that determine the equivalence between the minimal number of
gates in the standard circuit simulation and the length of the geodesic.
Both upper and lower bound are determined by the simulation parameters and,
of course, the length of the geodesic.

We consider the two cases. First: the simulation of the geodesic with set of
gates $\mathcal{G}$ that is exactly universal, and the case with
approximately universal set of gates. With the exactly universal set of
gates for any point\ $x_{0}\in \mathcal{U}$ there is a gate in a set $%
\mathcal{G}$ by which we can simulate exactly any point $x_{1}\in \mathcal{U}
$ in the ball of radius $\ r$ centered around $x_{0};$ \ we denote this ball
as $B_{x_{0}}^{+}(r)$. Under that assumption, we prove that the number of
gates in the simulation of a geodesic may be upper and lower bounded by a
linear factor in the length of geodesic and simulation parameters.

When the set of gates, $\mathcal{G}_{\mathcal{\epsilon }}$, is approximately
universal, a single gate from this set can simulate the points in $%
B_{x_{0}}^{+}(r)$ only with some finite precision $\epsilon $, and that will
necessarily mean that the circuit that simulates the geodesic is doing so
along the path that is not shorter than the actual geodesic, for that very
point. 

Our aim here is not to elaborate on the algorithm for the geodesic
simulation but rather to prove the bounds that optimal simulation can
achieve. We say optimal, because the set of gates $\mathcal{G}$ that we
first consider is much more powerful than any local and universal set of
gates. Therefore the result that we present is the\ optimal result about
complexity equivalence between discrete and continuous notions of
computation. In particular, for the standard simulation model described in 
\cite{Nielsen06}, we derive the highest lower bound and the lowest upper
bound in the minimal circuit complexity that one can hope to achieve with
the simulation of a geodesic.

\section{Preliminaries}

A quantum operation $U\in \mathcal{U}$ is a point on the manifold $\mathcal{U%
}\equiv SU(2^{n})$ at some distance from identity $I\in \mathcal{U}.$ The
distance considered is the integral distance that is determined by the
structure used on the manifold. In general that structure may be more
general than Riemannian, i.e. it is called the Finsler structure $F(x,y)$.
The restriction of a Finsler structure $F$ to any specific tangent space $%
T_{x}\mathcal{U}$ with the origin at the point $x\in \mathcal{U}$ is called
Minkowski norm on $T_{x}\mathcal{U}$. The second argument of the structure $%
F(x,y)$ is the velocity and its definition follows. Therefore a Finsler
structure is basically a family of the smoothly varying Minkowski norms, one
for each tangent space.

The defining properties of a non-negative real-valued structure $F(x,y)$ on $%
\mathbb{R}^{4^{n}-1}$ are as follows:

\begin{itemize}
\item[(1)] it is $C^{\infty }$ anywhere on $\mathbb{R}^{4^{n}-1}$ except at $%
y=0$;

\item[(2)] it is positive homogeneous, i.e. $F(x,\lambda y)=\lambda F(x,y)$
for $\lambda >0$;

\item[(3)] the $(4^{n}-1)$x$(4^{n}-1)$ matrix $\frac{\partial }{\partial
y_{i}}\frac{\partial }{\partial y_{j}}[\frac{1}{2}F^{2}]$ is positive
definite unless $y=0.$ As a consequence, one can derive positivity and
triangular equality of Minkowski norms \cite{Bao00}. The structure $F(x,y)$
is usually denoted simply as $F(y).$
\end{itemize}

For any $a,b\in \mathbb{R}_{+}$ we say that a map $\sigma :[a,b]\rightarrow 
\mathcal{U}$ is a piecewise $C^{\infty }$ curve with velocity $y\equiv \frac{%
d\sigma }{dt}=\sum_{i}\frac{d\sigma _{i}}{dt}\frac{\partial }{\partial x_{i}}%
\in T_{\sigma (t)}\mathcal{U}$. The integral length of the curve $\sigma ,$ $%
L(\sigma ),$ is defined as: 
\begin{equation}
L_{F}(\sigma )=\int_{a}^{b}F(\sigma ,\frac{d\sigma }{dt})dt\,\,.
\end{equation}%
Since we are usually interested in minimum length curves for $x_{0},x_{1}\in 
\mathcal{U}$, we denote by $\Gamma (x_{0},x_{1})$ the collection of all
piecewise $C^{\infty }$ curves $\sigma :[a,b]\rightarrow \mathcal{U}$ such
that $\sigma (a)=x_{0}$ and $\sigma (b)=x_{1}$. Similarly, the integral
distance is defined as a map $d_{F}:\mathcal{U}\times \mathcal{U}\rightarrow
\lbrack 0,\infty )$: 
\begin{equation}
d_{F}(x_{0},x_{1})=\inf_{\Gamma (x_{0},x_{1})}L_{F}(\sigma )
\end{equation}%
Using these definitions, one can show that the Finsler manifold $(\mathcal{U}%
,d_{F})$ satisfies the two axioms of a metric space: (1) positivity: $%
d_{F}(x_{0},x_{1})\geq 0$, where equality holds if and only if $x_{0}=x_{1}$
and (2) the triangular inequality: $d_{F}(x_{0},x_{2})\leq
d_{F}(x_{0},x_{1})+d_{F}(x_{1},x_{2})$. In general, the symmetric property
of a distance does not need to hold, and therefore $d_{F}(x_{0},x_{1})\neq
d_{F}(x_{1},x_{0})$.

\section{Distortion Lemma}

To establish the equivalence result, we introduce in this section the main
tool of our analysis. The intuitive idea on which we build our results
relies on the relation between the distances on the manifold and the
distances on the tangent space of the manifold. While the former are
introduced by the unitary gates and their complexity, the latter are defined
by the appropriately defined distances between the Hamiltonians of gates
used in the simulation. This will be proven useful in the sections below.

The lemma that follows is a slightly stronger result of a well-known and
very useful fact from the differential geometry. Again, it relates distances
on the manifold with the minimum and maximum distortion of \ the Euclidian
norm on the tangent space over the compact set. Interested reader are
encouraged to consult \cite{Bao00}, an excellent and very elaborate
reference on this subject.

\begin{lemma}[Distortion Lemma]
\textbf{\label{lemma CE}} Let $(\mathcal{U},F)$ be a Finsler manifold, and
for any point $x\in \mathcal{U}$ let $\varphi :P_{x}\rightarrow \mathbb{R}%
^{4^{n}-1}$ be the local coordinate system diffeomorphism of a compact set $%
P_{x}$ onto an open ball of $\mathbb{R}^{4^{n}-1}$, such that $\varphi (x)=0$%
. Then for a given $x_{0},\,\,x_{1}\in P_{x}$, and any Finsler metric \ $%
F(x,y),$ there exist a constant minimum $\mathfrak{m}>0$ and a constant
maximum $\mathfrak{M}>1$ such that the following relation is true:
\end{lemma}

\begin{equation}
\mathfrak{m}|\varphi (x_{1})-\varphi (x_{0})|\leq L_{F}(x_{0},x_{1})\leq 
\mathfrak{M}\,|\varphi (x_{1})-\varphi (x_{0})|\,\,.  \label{Lf}
\end{equation}%
Here $|\varphi (x_{1})-\varphi (x_{0})|\,\ $denotes the Euclidean length of
the $4^{n}-1$ dimensional vector in the tangent space.

\textbf{Proof: }We first note that a compact set $P_{x}$ for which $\varphi
(x)=0$ always exists. This is true because, given a local coordinate system $%
\varphi :Q\rightarrow \mathbb{R}^{4^{n}-1}$ and $x\in $ $Q$ for which $%
\varphi (x)=0,$ we can choose $P_{x}$ to be a closure of the preimage of $%
\varphi ^{-1}(B^{4^{n}-1}(r))$ for some $r>0.$ By $B^{4^{n}-1}(r)=\{v\in 
\mathbb{R}^{4^{n}-1}:|v|=\sqrt{\sum_{i}v_{i}^{2}}<r\}$ we denote the ball of
radius $r$ in the tangent space whose closure is a subset of $\varphi (Q).$

Next we note that, for tangent vector $y=\sum_{i}y_{i}\frac{\partial }{%
\partial x_{i}}\equiv \frac{dx}{dt}\in T_{x}\mathcal{U},$ the ratio between
Minkowski norm $F(x,y)$ and $x$-dependent Euclidean norm $|y|:=\sqrt{%
\sum_{i}y_{i}^{2}}$ for the basis $\{\frac{\partial }{\partial x_{i}}\}$ is
well defined for $y\neq 0.$ Since both norms are positive continuous
functions over the compact sets their quotient is also a positive continuous
function. Therefore the quotient's minimum $\mathfrak{m}$ and maximum $%
\mathfrak{M}$ exist and are both positive: $0<\mathfrak{m}\leq \frac{F(y)}{%
|y|}\leq \mathfrak{M}$. In other words, for all $y\in T_{x}\mathcal{U}$ and
all $x\in P_{x}$: 
\begin{equation}
\mathfrak{m}|y|\leq F(x,y)\leq \mathfrak{M}|y|\mathfrak{.}  \label{Cs}
\end{equation}

Now we can prove the right hand side (RHS) of inequality (\ref{Cs}) by
choosing the path $\sigma \in P$ \ that maps under $\varphi $ to a line
segment. In that case we can write: 
\begin{equation}
L_{F}(x_{0},x_{1})=\int_{t_{0}}^{t_{1}}F(\sigma ^{\prime })dt\leq \mathfrak{M%
}\int_{t_{0}}^{t_{1}}|\sigma ^{\prime }|dt=\mathfrak{M}\,|\varphi
(x_{1})-\varphi (x_{0})|\,\,,  \label{RHS}
\end{equation}%
where $\sigma ^{\prime }=\frac{d\sigma }{dt}$ denotes the velocity field of
a path $\sigma .$

To prove the left hand side of inequality (\ref{Cs}) we first show that $%
\sigma $ must be contained in $P_{x}$. The proof is by contradiction as
follows.

Choose $r_{0}<\frac{\mathfrak{m}}{\mathfrak{m+}3\mathfrak{M}}r$ and $%
\epsilon _{0}=\mathfrak{M}r_{0}$ and $P_{0}=\varphi
^{-1}[B^{n}(r_{0})]\subset P_{x}$. Let $\sigma :[t_{0},t_{1}]\rightarrow 
\mathcal{U}$ be a piecewise $C^{\infty }$ curve such that $\sigma
(t_{0})=x_{0}$ and $\sigma (t_{1})=x_{1}$ for $x_{0},\,x_{1}\in P_{0}$. If $%
L_{F}(\sigma )\leq d_{F}(x_{0},x_{1})+\epsilon _{0}$ then the curve $\sigma $
is certainly contained in $P_{x}$, and since by equation (\ref{RHS}) $%
d_{F}(x_{0},x_{1})\leq 2\mathfrak{M}r_{0}$ we have by assumption that $%
L_{F}(\sigma )\leq 3\mathfrak{M}r_{0}$. Now if we suppose that $\sigma $ is
not contained in $P_{x}$, and let $t_{0}\leq t^{\ast }\leq t_{1}$ be the
first instance where $\sigma $ reaches the boundary $\partial P_{x}$, at the
point $q\equiv \sigma (t^{\ast })$, so that $|\varphi (q)|=r$, then: 
\begin{equation}
L_{F}(\sigma )\geq L_{F}(\sigma _{\lbrack t_{0},t^{\ast
}]})=\int_{t_{0}}^{t^{\ast }}F(\sigma ^{\prime })dt\geq \mathfrak{m}%
\int_{t_{0}}^{t^{\ast }}|\sigma ^{\prime }|dt\geq \mathfrak{m}|\varphi
(q)-\varphi (x_{0})|\geq \mathfrak{m}(r-r_{0})\,\,.  \label{LHS}
\end{equation}

But the length of this curve would in fact be longer then the maximum
possible length of $3\mathfrak{M}r_{0}<\mathfrak{m}(r-r_{0}),$ since by
assumption we are assured that for $\mathfrak{m}>0$ and $\mathfrak{M}>1$ it
is true that $r_{0}\,<\frac{\mathfrak{m}}{\mathfrak{m+}3\mathfrak{M}}r.$
Therefore $\sigma $ must be contained in $P_{x}$. \newline

The proof of the left hand side of inequality (\ref{Lf}) follows by the same
arguments as were used to prove (\ref{LHS})$\square $

Given the distortion lemma for the length of the path for any two points
that belong to the compact set, we can easily derive a simillar result that
is valid for the shortest distances.

\begin{corollary}
\textbf{\label{corollary CE}}For a Finsler manifold $(\mathcal{U},F)$, and
any point $x\in \mathcal{U}$, let $\varphi :P_{x}\rightarrow \mathbb{R}%
^{4^{n}-1}$ be the local coordinate system diffeomorphism of a compact set $%
P_{x}$ onto an open ball of $\mathbb{R}^{4^{n}-1}$, such that $\varphi (x)=0$%
. Then, for a given $x_{0},\,\,x_{1}\in P_{x}$ and any Finsler metric $%
F(x,y) $ there exist a constant minimum $\mathfrak{m}>0$ and a constant
maximum $\mathfrak{M}>1$ such that the following relation is true:
\end{corollary}

\begin{equation}
\mathfrak{m}|\varphi (x_{1})-\varphi (x_{0})|\leq d_{F}(x_{0},x_{1})\leq 
\mathfrak{M}\,|\varphi (x_{1})-\varphi (x_{0})|\,\,.
\end{equation}

\textbf{Proof: }We only need to verify the left hand side of inequality (\ref%
{Lf}) is still true for minimal length curves. By definition of metric
distance, for $0\leq \epsilon \leq \epsilon _{0}$, two points $%
x_{0},x_{1}\in P_{0}$ can be joined by a piecewise $C^{\infty }$ curve $%
\sigma :[t_{0},t_{1}]\rightarrow \mathcal{U}$ with integral length: 
\begin{equation}
L_{F}(\sigma )\leq d_{F}(x_{0},x_{1})+\epsilon \,\,.
\end{equation}%
By previous arguments, $\sigma $ must lie in $P_{x}$, end by similar
calculations we find that: 
\begin{equation}
\mathfrak{m}|\varphi (x_{1})-\varphi (x_{0})|\leq L_{F}(\sigma )\leq
d_{F}(x_{0},x_{1})+\epsilon \,\,,\text{ }
\end{equation}%
Letting $\epsilon \rightarrow 0$ proves the desired result. $\square $

Lemma (\ref{lemma CE}) and Corollary (\ref{corollary CE}) allow us to bound
the lengths on the manifold to the Euclidian lengths on the tangent space.
For Euclidean coordinates in our tangent space we will have the coefficients
in the decomposition of the gate Hamiltonian matrix in terms of the
generalized Pauli matrices, $n$ times tensored two dimensional matrices from
the set $\{I,X,Y,Z\}.$

For example, in the context of simulation, points $x_{0},\,x_{1}\in P_{x}$
and an open set \thinspace $P_{x}$ are chosen such that they correspond to
the end points in the simulation by a single gate. Moreover, we can
construct a \emph{local coordinate system} on the Lie group $SU(2^{n})$
which is a Lie algebra $\mathfrak{su}(2^{n})$. For the \emph{origin} $%
x_{s}\in \mathcal{U}$, define a pull back map $\varphi ^{-1}:\mathbb{R}%
^{4^{n}-1}\rightarrow \mathcal{U}$, so that $x_{s+1}\equiv \exp
^{-iy_{s+1}\cdot \sigma }x_{s}=\exp ^{-i\varphi (x_{s+1})\cdot \sigma }x_{s}$%
, where $\sigma $ denotes the coordinate basis, i.e. $(4^{n}-1)$-component
vector whose entries are the generalized Pauli matrices.

For the particularly chosen metric, as in \cite{Nielsen06}, $%
F_{p}(x_{0},y)\equiv F_{p}(y)=\sqrt{\sum_{i=1}^{k}y_{i}^{2}+p^{2}\sum_{j\neq
i}y_{i}^{2}},$ where $k=\frac{9(n^{2}-n)}{2}+3n$, which introduces a penalty 
$p$ for the subset of Hamiltonian coordinates in the tangent space, so we
have:

\begin{equation*}
|y|\leq F_{p}(y)\leq p|y|.
\end{equation*}%
Since this relation is true on any compact set, by Corollary (\ref{corollary
CE}) we have:

\begin{equation}
|\varphi (x_{s+1})-\varphi (x_{s})|\leq d_{F_{p}}(x_{s},x_{s+1})\leq
p|\varphi (x_{s+1})-\varphi (x_{s})|\,.  \label{nm}
\end{equation}%
It is important to note that in our analysis constants $\mathfrak{m}$ and $\ 
\mathfrak{M}$ do not depend on the compact set within which each gate is
applied, and they are basically the property of the metric. This property
might not be true in general for some other Finsler structures, but for our
purposes here this assumption is very plausible.

\section{Equivalence Result}

For the sake of consistency and easier understanding, we follow the notation
from \cite{Nielsen05} and denote with $m_{\mathcal{G}}$ the minimum number
of gates, for a given set\ $\mathcal{G},$ needed to implement an arbitrary
unitary $U\in \mathcal{U}$. Moreover, in this section we assume that the
geodesic is simulated by sequential application of the gates from the set $%
\mathcal{G},$ and that\ by using a single gate from the set $\mathcal{G}$ we
can simulate exactly any other point in the ball $B_{x_{0}}^{+}(\epsilon ),$
i.e. which is the $\epsilon -$neighborhood around the initial condition at
the point $x_{0}$. This is an unrealistic scenario, since the set of gates
would need to be infinite and non-local. Hence, we relax it in the next
section. However, for the purpose of exact simulation of the geodesic it is
an important tool. Clearly consideration of the set of gates defined in this
way, as we shall see, is the best we can possibly hope for, and thus the
bounds achieved by the set $\mathcal{G}$ \ are optimal, i.e. they determine
the bounds achievable by any other set of gates that is less powerful.

For any gate used in a simulation we assign a gate index, so that eventually
the index set is $s=\{0,1,2,...,m_{\mathcal{G}}-1\}$ for every gate in the
simulation. Moreover, by $\sigma (t):[0,m_{\mathcal{G}}]\rightarrow \mathcal{%
U}$ we denote a minimal geodesic between $\sigma (0)=I,\,\sigma (m_{\mathcal{%
G}})=U$.

Note that, since $\mathcal{U}$ is a compact manifold with Finsler structure,
all forward and backward Cauchy sequences with respect to $d$ must converge
on \QTR{cal}{U}. More precisely, compact Finsler spaces are automatically
both forward complete and backward complete. This fact holds regardless of
whether the Finsler structure is absolutely homogeneous or only positively
homogeneous. Therefore, any two points on the manifold can be connected by a
minimizing geodesic, as that property itself is a sufficient condition for
the Hopf-Rinow theorem \cite{Bao00}.

\begin{theorem}
\textbf{\label{thm: Main}}Let $d_{F}(I,U)$ denote a length of a geodesic
between $I$ and $U\in SU(2^{n}).$ For any simulation index set $%
s=\{0,1,2,...,m_{\mathcal{G}}-1\}$ let $P_{x_{s}}\in \mathcal{U}$ be an open
set on the manifold that contains a segment of minimizing geodesic $\sigma
_{s}(t):[s,s+1]\rightarrow \mathcal{U},$ that is simulated exactly by a
single gate. Moreover, let $P_{x_{s}}$ be mapped by $\varphi $
diffeomorphically onto an open ball in $\mathbb{R}^{4^{n}-1},$ so that $\rho
_{s}=$ $|\varphi (x_{s+1})-\varphi (x_{s})|$ is the Euclidean length of the
image of the geodesic segment $\sigma _{s}$. If we denote $\rho _{\sup
}=\sup_{s}\rho _{s}$ and $\rho _{\inf }=\inf_{s}\rho _{s},$ then the
following relation holds:%
\begin{equation}
\frac{d_{F}(I,U)}{\rho _{\sup }\mathfrak{M}}\,\leq m_{\mathcal{G}}\leq \frac{%
d_{F}(I,U)}{\rho _{\inf }\mathfrak{m}}\,\,.  \label{main}
\end{equation}
\end{theorem}

\textbf{Proof:} For any segment gate index from set $s,$ by the Corollary (%
\ref{corollary CE}) we see that:

\begin{equation*}
\mathfrak{m}\rho _{s}\leq d_{F}(x_{s},x_{s+1})\leq \mathfrak{M}\rho _{s}\,,
\end{equation*}

Summing over all segments of minimizing geodesic $%
\sum_{s=0}^{m_{g}-1}d_{F}(x_{s},x_{s+1})=d_{F}(I,U)$, and taking into
account that $\sum_{s=0}^{m_{g}-1}\beta _{s}\leq m_{\mathcal{G}}\beta $ and $%
m_{\mathcal{G}}\frac{\rho ^{2}}{\beta }\leq \sum_{s=0}^{m_{g}-1}\frac{\rho
_{s}^{2}}{\beta _{s}},$ it is easy to see that: 
\begin{equation}
m_{\mathcal{G}}\mathfrak{m}\rho _{\inf }\leq
d_{F}(I,U)=\sum_{s=0}^{m_{g}-1}d_{F}(x_{s},x_{s+1})\leq m_{\mathcal{G}}%
\mathfrak{M}\rho _{\sup }\,\,,
\end{equation}%
which gives the desired result by rearranging the variables.$\square $

Note that the above theorem is derived in terms of bounds of the Euclidean
distances in the tangent space. One may take a different path though, as for
example Nielsen in \cite{Nielsen05}, by deriving the result for the lower
bound in terms of the lengths of the geodesic segments simulated by the
single gate: $d_{F}(x_{s},x_{s+1})\leq \beta _{\sup }.$ From the following
theorem, one can reproduce the result derived by Nielsen as a special case
when $\beta _{\sup }=1.$

\begin{theorem}
\textbf{\label{thm: Main 1}}Let $d_{F}(I,U)$ denote a length of a geodesic
between $I$ and $U\in SU(2^{n}).$ For any simulation index set $%
s=\{0,1,2,...,m_{\mathcal{G}}-1\}$ let $P_{x_{s}}\in \mathcal{U}$ be an open
set on the manifold that contains a segment of minimizing geodesic $\sigma
_{s}(t):[s,\,s+1]\rightarrow \mathcal{U},$ that is simulated exactly by a
single gate. Moreover, let $P_{x_{s}}$ be mapped by $\varphi $
diffeomorphically onto an open ball in $\mathbb{R}^{4^{n}-1},$ so that $\rho
_{s}=$ $|\varphi (x_{s+1})-\varphi (x_{s})|$ is an image of the bounded
length geodesic segment $d_{F}(x_{s},x_{s+1})\leq \beta _{s}$. If we denote $%
\beta _{\sup }=\sup_{s}\beta _{s}$ and $\beta _{\inf }=\inf_{s}\beta _{s},$
then the following relation holds:%
\begin{equation}
\frac{d_{F}(I,U)}{\beta _{\sup }}\leq m_{\mathcal{G}}\leq \frac{\mathfrak{M}%
}{\mathfrak{m}}\frac{d_{F}(I,U)}{\beta _{\inf }}\ .
\end{equation}
\end{theorem}

\textbf{Proof:} Following along the lines of Theorem (\ref{thm: Main}):

\begin{equation*}
\mathfrak{m}\frac{\beta _{s}}{\mathfrak{M}}=\mathfrak{m}\rho _{s}\leq
d_{F}(x_{s},x_{s+1})\leq \beta _{s}\equiv \mathfrak{M}\rho _{s}\,,
\end{equation*}

Summing over all segments of minimizing geodesic $%
\sum_{s=0}^{m_{g}-1}d_{F}(x_{s},x_{s+1})=d_{F}(I,U)$, and taking into
account that $m_{\mathcal{G}}\beta _{\inf }\leq \sum_{s=0}^{m_{g}-1}\beta
_{s}\leq m_{\mathcal{G}}\beta _{\sup }$: 
\begin{equation}
\frac{\mathfrak{m}}{\mathfrak{M}}m_{\mathcal{G}}\beta _{\inf }\leq
d_{F}(I,U)\leq m_{\mathcal{G}}\beta _{\sup }\,\,,  \label{main 1}
\end{equation}%
which gives the stated result. $\square $

Equations (\ref{main}) and (\ref{main 1}) establish the tightest possible
equivalence between the minimal number of gates in the circuit and geodesic
length as a function of the simulation parameters. Again, the simulation\
parameters may be defined in terms of distances traversed with the single
gate on the manifold or in terms of the Euclidean distances between the
initial and final coefficients in the generalized Pauli expansion of the
gate Hamiltonian. Even though the above results give no indication as to how
to implement the simulation, they do provide us the best bounds we currently
have and give us an estimate to the quality of the simulation provided that
one knows the simulation parameters. However, the above results can be
applied to the arbitrary paths on the manifold including those that are
generated adiabatically. In particular, it would be very interesting to
compare the results for bounds of circuit size obtained by geometric
techniques with the equivalence results obtained in \cite{Aharonov04}.

\section{Approximate simulation}

In this section we reformulate the bounds for the standard circuit
simulation procedure where the set of gates used consists solely of the
single and two qubit gates, which are applied sequentially. Since the exact
simulation of arbitrary unitary gate by single and two qubit gates demands
an exponential number of gates, almost all unitaries simulated by the
polynomial number of gates will be simulated approximately.

In particular, we consider two paths. Let the first be $d_{F_{p}}(I,\tilde{U}%
)$, denoting the length of the geodesic simulated exactly with the set of
gates from $\mathcal{G}$ with respect to the Finsler metric $F_{p}$ , and
let the second one $L_{F_{p}}(I,\tilde{U})$ be the minimum length path for
the exact simulation of $\tilde{U}$ by the set of gates from $\mathcal{G}%
_{2}.$ Here we denote by $\mathcal{G}_{2}$ the set of unitary gates whose
time independent Hamiltonians have Hamming weight not greater than two.

Note that the length $L_{F_{p}}(I,\tilde{U})$ has nothing to do with $%
d_{F_{p}}(I,\tilde{U})\,,\,$\ as $L_{F_{p}}(I,\tilde{U})$ is completely
determined by the simulation, and almost everywhere does not simulate the
geodesic $d_{F_{p}}(I,\tilde{U})\,.$

\begin{corollary}
\textbf{\label{corollary CE1}}Let $\tilde{U}$ be the approximation of the
unitary operation $U$ that is simulated by the one and two qubit gates. Then
the lower bound on the minimum circuit size $\tilde{m}_{\mathcal{G}_{2}}$ is
at most $O(\frac{n^{4}}{p}d_{F_{p}}^{2}(I,U)L_{F_{p}}(I,\tilde{U})),$ and
the upper bound on $m_{\mathcal{G}_{2}}$ is at least $\Omega
(n^{4}d_{F_{p}}^{3}(I,U))\,\,\,.$
\end{corollary}

\textbf{Proof:} The three step standard simulation of arbitrary $%
U=e^{-iH(t)t}$ is elaborated in detail by Nielsen, Dowling, Gu and Doherty
in \cite{Nielsen06}. The procedure can be sketched as follows:

\begin{itemize}
\item[(1)] the time variable Hamiltonian $H(t)$ is substituted by projected
the Hamiltonian $H_{P}(t)$ that is formed by deleting all $\sigma _{i}$ for $%
i>k,$ i.e. all three- and more-body terms in the Pauli expansion of $%
H(t)=\sum_{i=1}^{k}y_{i}\sigma _{i}+\sum_{j\neq i}y_{i}\sigma _{i},$ where $%
k=\frac{9(n^{2}-n)}{2}+3n;$

\item[(2)] the evolution due to $H_{p}(t)$ is broken up into many small
intervals, each of length $\Delta ,$ over which the time-dependent
Hamiltonian $H_{p}(t)$ is accurately simulated by a constant mean
Hamiltonian $\bar{H}_{p}^{\Delta }$;

\item[(3)] the mean Hamiltonian $\bar{H}_{p}^{\Delta }$ that has $k$ terms
in the Pauli expansion with coefficients $|y_{i}|\leq 1$ is simulated with a
standard simulation technique \cite{Chuang00} using one and two qubit gates.
\end{itemize}

The reader is encouraged to see \cite{Nielsen06} for full detail of the
approximation result.

For the above procedure, since $SU(2^{n})$ is compact and simply connected,
there exists a path $L_{F_{p}}(I,\tilde{U})$ that is exactly synthesized
with the gates in the simulation. By exactly simulated we mean that the end
points of each gate in the simulation lie precisely on the path of length $%
L_{F_{p}}(I,\tilde{U}).$ Clearly, the length of $L_{F_{p}}(I,\tilde{U})\geq $
$d_{F_{p}}(I,U).$

Now we bound length of the path segments, $L_{F_{p}}(\tilde{x}_{s},\tilde{x}%
_{s+1}),\,\ $for each of $\tilde{m}_{\mathcal{G}_{2}}$ gates in the
simulation. Since there exists a compact set $P_{x_{s}},$ such that end
points $\tilde{x}_{s},\tilde{x}_{s+1}\in $ $P_{x_{s}},$ that maps
diffeomorphically to the local coordinate system, we can use Lemma (\ref%
{lemma CE}) and its corollaries. Corroborating the arguments used to derive
equation (\ref{nm}), over the compact set $P_{x_{s}}$, \ the Finsler
structure, i.e. the Minkowski norm for the Pauli expansion of $H(t)$, is $%
F_{p}(x_{s},y_{s})=\sqrt{\sum_{i=1}^{k}y_{i}^{2}+p^{2}\sum_{j\neq i}y_{i}^{2}%
}$. Its minimum and maximum distortion over the compact set $P_{x_{s}}$ are: 
$|y_{s}|\leq F_{p}(y_{s})\leq p|y_{s}|.$ Therefore, by the Lemma (\ref{lemma
CE})

$|\varphi (\tilde{x}_{s+1})-\varphi (\tilde{x}_{s})|\leq L_{F_{p}}(\tilde{x}%
_{s},\tilde{x}_{s+1})\leq p|\varphi (\tilde{x}_{s+1})-\varphi (\tilde{x}%
_{s})|.$

The same is true for any other segment in the simulation, and hence:%
\begin{equation}
m_{\mathcal{G}_{2}}\rho _{\inf }^{\Delta }\leq L_{F_{p}}(I,\tilde{U}%
)=\sum_{s=0}^{m_{g}-1}L_{F_{p}}(\tilde{x}_{s},\tilde{x}_{s+1})\leq m_{%
\mathcal{G}_{2}}p\rho _{\sup }^{\Delta }  \label{sim}
\end{equation}%
where $\rho _{\inf }^{\Delta }=\inf_{s}$ $|\varphi (\tilde{x}_{s+1})-\varphi
(\tilde{x}_{s})|,$ and $\rho _{\sup }^{\Delta }=\sup_{s}$ $|\varphi (\tilde{x%
}_{s+1})-\varphi (\tilde{x}_{s})|$. Note that we can always choose the $s$%
-th gate local coordinate system so that $\varphi (\tilde{x}_{s})=0.$

Finally, in the three-step simulation summarized above, gates at the third
stage simulate the time invariant Hamiltonian $\bar{H}_{p}^{\Delta }$ for
the segment $\Delta ,$ with coordinates $|y_{i}|\leq 1.$ More precisely, the 
$s$-th gate simulates the neighborhood $\ $around $x_{s}$: $x_{s+1}\equiv
e^{-i\varphi (x_{s+1})\cdot \sigma }x_{s}=e^{-iy_{s}\sigma _{s}\Delta
^{2}}x_{s}.$ Here $\sigma \in $ $\mathcal{G}$ denotes stabilizer basis on $n$
qubits, $\sigma _{s}\in $ $\mathcal{G}_{2}$, and $\Delta ^{2}$ is the
simulation time for every gate. If we choose $\Delta =\Theta
((n^{2}d_{F_{p}}(I,U))^{-1}),$ as in \cite{Nielsen06}, then for $|y_{s}|\leq
1$ we see that $\rho _{s}=$ $\left\vert y_{s}\Delta ^{2}\right\vert =\Theta
((n^{4}d_{F_{p}}^{2}(I,U))^{-1})$. Finally, using equation (\ref{sim}) we
establish that the simulation with $\tilde{m}_{\mathcal{G}_{2}}$ gates has
the upper bound $\Theta (n^{4}d_{F_{p}}^{2}(I,U)L_{F_{p}}(I,\tilde{U}))\geq
\Omega (n^{4}d_{F_{p}}^{3}(I,U)).$ By similar arguments, for the lower
bound, we get $O(\frac{n^{4}}{p}d_{F_{p}}^{2}(I,U)L_{F_{p}}(I,\tilde{U}))$. $%
\square $

\section{Conclusion}

The Distortion Lemma and its corollary provide a general tool for relating
distances on the manifold with distances on the tangent space. In this paper
we have derived a generalized linear bounds for the exact simulation of any
path on the manifold, in terms of the minimum circuit size and the
simulation parameters.

The equivalence between the path on the manifold and circuit size still
persists in the case of approximate simulation, provided that the simulation
parameters have the appropriate scaling. However, one can not expect better
than $n^{2}$ times improvement in the minimum circuit size upper bound over
the result for standard circuit simulation derived by Nielsen, Dowling, Gu
and Doherty \cite{Nielsen06}.

Moreover, if\ one defines a metric on the manifold that penalizes the
hard-to-simulate directions on the tangent space with high cost, that cost\
is, in effect, translated to the increased ratio between upper and lower
bound in minimum circuit size.

\section{Acknowledgements}

M.D. acknowledges NSF for its support under the ITR Grant No. EIA-0205641,
and to Umesh Vazirani, Michael Huchings and Ben Reichardt for useful
discussions.

\end{document}